\documentclass[12pt]{article}
\usepackage{geometry}                
\usepackage{graphicx}
\usepackage{amssymb}
\usepackage{epstopdf}
\def\mass{\mathcal M}
\def\hor{\cal H}

\def\gzero{g^{(0)}_{ab} }

\def\szero{s^{(0)}_{ab} }
\def\pizero{\pi _{(0)}^{ab} }
\def\lzero{\Lambda _{(0)} }
\def\tl{\tilde{\Lambda }}
\def\fzero{f_0}
\def\gzero{g^{(0)} _{ab}}

\def\ei{e _{(i)} }

\def\wab{\omega ^{ab}}

\begin{document}

\begin{titlepage}
\vfill
\begin{flushright}
\end{flushright}

\vfill
\begin{center}
\baselineskip=16pt
{\Large\bf Enthalpy and the Mechanics of AdS Black Holes}
\vskip 1.0cm
{\large {\sl }}
\vskip 10.mm
{\bf David Kastor${}^{a1}$, Sourya Ray${}^{b2}$ and Jennie Traschen${}^{a3}$} \\
\vskip 1cm
{
	${}^a$ Department of Physics, University of Massachusetts, Amherst, MA 01003\\	
     	${}^b$ Centro de Estudios Cientõficos (CECS), Casilla 1469, Valdivia, Chile \\
	${}^1$ \texttt{kastor@physics.umass.edu,} ${}^2$ \texttt{ray@cecs.cl,} ${}^3$ \texttt{traschen@physics.umass.edu}
     }
\vspace{6pt}
\end{center}
\vskip 0.5in
\par
\begin{center}
{\bf Abstract}
 \end{center}
\begin{quote}
We present geometric derivations of the Smarr formula for static AdS black holes and an expanded first law that includes variations in the cosmological constant.  These two results are further related by a scaling argument based on Euler's theorem.
The key new ingredient in the constructions is a two-form potential for the  static Killing field.
Surface integrals of the Killing potential determine the coefficient of  the variation of $\Lambda$ in the first law.  This coefficient is proportional to a finite, effective volume for the region outside the AdS black hole horizon, which can also be interpreted
as minus the volume excluded from a spatial slice by the black hole horizon.  
This  effective volume also contributes to the Smarr formula.
Since $\Lambda$ is naturally thought of as a pressure, the new term in the first law has the form of 
effective volume times change in pressure that arises in the variation of the enthalpy in classical thermodynamics.
This and related arguments suggest that the mass of an  AdS black hole should be interpreted as the enthalpy of the spacetime.
\vfill
\vskip 2.mm
\end{quote}
\end{titlepage}


\section{Introduction}

Black hole solutions with a non-vanishing cosmological constant have received considerable recent attention.   This is due both to the role they play in the phenomenology of the AdS/CFT correspondence  \cite{Maldacena:1997re,Gubser:1998bc,Witten:1998qj} and also, of course, to the observational data suggesting that the universe may  have a small, positive value of $\Lambda$ (see {\it e.g. }reference \cite{Spergel:2006hy}).  In four dimensions, the extension of the Kerr-Newman family of solutions to non-zero $\Lambda$ was found many  years ago by Carter \cite{Carter:1968ks}.  More recently 
amongst many other results, the extension of the Myers-Perry higher dimensional rotating black hole solutions 
\cite{Myers:1986un} were found in
\cite{Hawking:1998kw,Gibbons:2004uw},  solutions including NUT charge were constructed in  \cite{Chen:2006xh, Chen:2006ea} and approximate black ring solutions in AdS were been studied in \cite{Caldarelli:2008pz}.  

The purpose of this paper is to bring our understanding of certain properties of AdS black holes more closely in parallel with well known results in the asymptotically flat case\footnote{Our focus will be on the AdS case, but our methodology applies to deSitter black holes as well.}. Specifically, we will  focus on the Smarr formula for AdS black holes  and on an associated, extended version of the first law  that accounts for variations in the black hole mass with respect to variations in the cosmological constant.  In the asymptotically flat case, the Smarr formula and first law can be derived by geometric means, without relying on the explicit solutions to the field equations.   Moreover, these two results are  related via a simple scaling argument.  We will show that similar geometric methods can be used to obtain the Smarr formula and first law with $\Lambda\neq 0$, and that these results are again related by a simple scaling argument.

In order to extend the geometrical contructions of the Smarr formula and first law to the $\Lambda\neq 0$ case, we employ a new technical ingredient, the Killing potential $\omega^{ab}$ that is related to a Killing vector $\xi^a$  according to 
\begin{equation}\label{kpot}
\xi^b=\nabla_a\omega^{ab}.
\end{equation}
The Killing potential was introduced in references  \cite{Bazanski:1990qd,Kastor:2008xb} in order to construct a Komar integral relation for $\Lambda\neq 0$.  We use this new Komar relation to derive the Smarr formula for static AdS black holes\footnote{For simplicity, we will restrict our considerations in this paper to static black holes.  However, results for the more general stationary case may be derived in  similar fashion.}.
We will see that the Killing potential can also be used in the context of the Hamiltonian perturbation theory techniques of 
references \cite{Traschen:1984bp}\cite{Sudarsky:1992ty}\cite{Traschen:2001pb} in order to prove the required generalization of the first law.

In fact, similar expressions for the  (A)dS Smarr formula  and the first law with 
variable $\Lambda$ have been obtained before in references 
\cite{Caldarelli:1999xj,Wang:2006eb,Sekiwa:2006qj,Wang:2006bn,Larranaga Rubio:2007jz,Cardoso:2008gm}.  The results in these papers 
have been established by direct calculation
for specific (A)dS black hole solutions, such as the BTZ black hole or Schwarzschild-(A)dS spacetimes.  
Our more general, geometrical approach, however, holds a number of advantages\footnote{Very recently Urano et. al.  \cite{Urano:2009xn} derived a first law with variable $\Lambda$ for Schwarzschild-dS spacetimes by means of an adaptation of the Iyer-Wald Noether charge formalism \cite{Wald:1993nt,Iyer:1994ys}.  This method is essentially a covariant version of the Hamiltonian perturbation theory which we use in Section (\ref{firstlawsection}).  However, the first law constructed in \cite{Urano:2009xn} differs substantially from that which would follow from application of our methods to deSitter black holes.  The resulting formula would relate variations in the black hole and deSitter horizon areas to the variation in the cosmological constant, while the first law in \cite{Urano:2009xn} relates the variation in the Schwarzschild-dS mass parameter to the variation in total sum of the horizon areas and the variation in $\Lambda$.}.  
Most importantly, we obtain a general expression for the quantity 
\begin{equation}\label{thetadef}
\Theta= 8\pi G(\partial M/\partial\Lambda)
\end{equation}
that appears in both the first law and the Smarr formula in terms of surface integrals of the Killing potential.  This is comparable to knowing that the quantity 
$\partial M/\partial A = \kappa/8\pi G$, where $\kappa$ is the black hole horizon surface gravity,
rather than having only an explicit expression given in terms of the parameters of a particular family of solutions (such as 
$\partial M/\partial A = 1/16\pi GM$ for Schwarzschild in $D=4$).  In addition,  our derivation of the first law with varying $\Lambda$, based on the approach of \cite{Sudarsky:1992ty}, holds for any perturbations that solve the linearized equations of motion around the static black hole background, rather than only for those that stay within the family of static solutions.

The new term in the first law has the form $\Theta\delta\Lambda/8\pi G$, 
with the quantity $\Theta$ having the dimensions of volume.  We show below that for AdS black holes, $\Theta$ is an effective renormalized volume $V$ for the region outside the black hole horizon, given by the difference between the (infinite) volume outside the horizon minus the (infinite) volume of a spatial slice of AdS.  The quantity $\Theta$ is then finite and negative.  It can also be thought of as minus the volume excluded by the black hole horizon.  If we think of the cosmological constant as a pressure $P$, then the new term in the first law (\ref{finalfirstlaw}) looks like $V\delta P$,  a term that occurs in the variation of the enthalpy $H=E +PV$ of a thermodynamic system.  This suggests that after expanding the set of thermodynamic variables to include the cosmological constant, the mass $M$ of an AdS black hole should be interpreted as the analogue of the enthalpy from classical thermodynamics, rather than as the total  energy of the spacetime.   This observation is the basis for the title of the paper.

Inclusion of the new term  in the first law  is motivated  in this paper  by the formal scaling argument that leads from the first law to the Smarr formula.  However, as noted in the references above, a number of physical mechanisms have been put forward for variable $\Lambda$.  Prominent among these is the model of Brown and Teitelboim \cite{Brown:1987dd,Brown:1988kg} in which the four dimensional cosmological constant represents the energy density of a $4$-form gauge field strength, which can change via the  instanton induced nucleation of charged  membranes.

We would also like to suggest that the inclusion of $\Lambda$ as a thermodynamic 
variable may find application in the AdS/CFT  
\cite{Maldacena:1997re,Gubser:1998bc,Witten:1998qj} context as well\footnote{See reference \cite{Papadimitriou:2005ii} for a treatment of the Smarr formula and first law with fixed $\Lambda$ in asymptotically locally AdS spacetimes from an AdS/CFT point of view.}.  
In its most well studied instance, AdS/CFT postulates an equivalence between type IIB string theory on  $AdS_5\times S^5$ and maximally supersymmetric $SU(N)$ Yang-Mills theory in four dimensions, where the radius of curvature $R = \sqrt {-6/\Lambda}$ of  $AdS_5$ (and the $S^5$) is given by 
$(R/l_s)^4 = g_{YM}^2 N$, with $g_{YM}$ the Yang-Mills coupling constant and $l_s$ is the string length.  The string coupling constant $g_s$ is given by $g_s= 4\pi g_{YM}^2$.  Supergravity is a good approximation in the regime $g_s\ll 1$ and $R/l_s\gg 1$ which requires $N$ to be large.  Assuming that the string coupling is held fixed, variation with respect to $\Lambda$ on the supergravity side translates into the variation with respect to number of colors $N$ on the field theory side.  It would be interesting to understand what the effective volume $\Theta$ corresponds to in the CFT.

Finally, our results in this paper for Einstein gravity with $\Lambda\neq 0$ provide the simplest example of  more general set of results that can be anticipated for Lovelock gravity theories \cite{Lovelock:1971yv}.   A first law for black holes in Lovelock gravity was derived in \cite{Jacobson:1993xs}.  However, a basic scaling argument suggests that the Smarr formula for a given Lovelock theory should be related to an augmented first law in which all the coupling constants but one are taken to vary.
The Komar integral relations for Lovelock gravity were found in \cite{Kastor:2008xb} and contain, at each higher derivative Lovelock order, a new quantity analogous to the Killing potential $\omega^{ab}$.  These should lead to the required Smarr formula and augmented first law for general Lovelock gravity theories.  The case we have considered in this paper is the one in which, rather than adding higher derivative Lovelock terms, we have added only the ``lower derivative" cosmological constant term to the Einstein action.

The paper proceeds in the following way.  In Section (\ref{scaling}) we give the scaling argument based on Euler's theorem that motivates consideration of a first law with varying $\Lambda$.  In Section (\ref{smarrsection}) we derive the Smarr formula for AdS black holes starting from the Komar integral relation for Einstein gravity with $\Lambda\neq 0$.  In Section (\ref{firstlawsection}) we derive the first law with variable $\Lambda$.  In Section (\ref{enthalpy}) we interpret the augmented first law as the variation in the enthalpy.  In Section (\ref{killingpotential}) we present a number of results relating to existence of the Killing potential in the near horizon region.

\section{Scaling argument}\label{scaling}

Let us begin with the scaling argument.  In the asymptotically flat case, Euler's theorem for homogeneous functions provides a route between the the first law of black hole mechanics and the Smarr formula for stationary black holes  (see {\it e.g.} the discussions in \cite{Gauntlett:1998fz,Townsend:2001rg}).    It was noted in \cite{Caldarelli:1999xj} (see also \cite{Gibbons:2004uw}) that in order to apply this procedure to the $\Lambda\neq 0$ case, one must  take into account the scaling properties of the cosmological constant.  This can be done in the following way.  From the gravitational action
$S = {1\over 8\pi G}\int d^Dx \sqrt{-g}\left( R - 2\Lambda\right)$,
we see that the 
cosmological constant  has dimension $(length)^{-2}$.  
Euler's theorem states that if a function $f(x,y)$ obeys the scaling relation $f(\alpha^p x, \alpha^q y) = \alpha^r f(x,y)$,
then the function and its partial derivatives satisfy the relation
\begin{equation}
r f(x,y)  = p\, \left({\partial f\over \partial x}\right)\,  x+ q\, \left({\partial f\over\partial y}\right)\, y.
\end{equation}
The mass $M$ of a static AdS black hole can be regarded as a function of the horizon area $A$ and the cosmological constant $\Lambda$.  Under an overall change in length scale, these quantities scale as $M\propto l^{D-3}$, $A\propto l^{D-2}$ and $\Lambda\propto l^{-2}$.
Euler's theorem then implies that the mass of a static AdS black hole satisfies the Smarr formula
\begin{equation}\label{smarrform}
(D-3)M = (D-2)\left({\partial M\over\partial A}\right) \, A - 2\left({\partial M\over\partial \Lambda}\right) \, \Lambda .
\end{equation}
Setting $\Lambda=0$ and  $\partial M/\partial A = \kappa/8\pi G$, from the first law of black hole mechanics, gives the well known Smarr formula for static, asymptotically flat black holes.  The partial derivative $\partial M/\partial\Lambda$   in (\ref{smarrform}) can similarly be found by considering an extended version of the first law for AdS black holes that includes the effect of varying the cosmological constant
\begin{equation}\label{firstlawform}
dM = 
{\kappa\over 8\pi G}\, dA + \left({\partial M\over\partial \Lambda}\right) \,  d\Lambda.
\end{equation}
The output of Sections (\ref{smarrsection}) and (\ref{firstlawsection}) can be thought of as two ways of determining this quantity 
$\partial M/\partial\Lambda$ by two different geometrical means.

\section{AdS Smarr formula}\label{smarrsection}

\subsection{Komar integral relation with $\Lambda\neq 0$}

The Smarr formula for asymptotically flat, stationary black holes can be found directly by geometrical means, {\it i.e.}  without making use of Euler's theorem.  The basic ingredient in the construction is the Komar integral relation 
\cite{Komar:1958wp}, a Gauss' law type statement that holds in a spacetime with a Killing vector.  Assume 
that $M$ is a spacetime satisfying the vacuum Einstein equations, that $\xi^a$ is a Killing vector on $M$ and that  
$\Sigma$ is a hypersurface in $M$ with boundary $\partial\Sigma$.  The associated Komar integral relation is the statement
\begin{equation}\label{komarrelation}
{D-2\over 8\pi G}\int_{\partial\Sigma}dS_{ab}\nabla^a\xi^b = 0,
\end{equation}
where $dS_{ab}$ is the volume element normal to the co-dimension $2$ surface $\partial\Sigma$, $D$ is the spacetime dimension and the prefactor is chosen for convenience\footnote{The volume element $dS_{ab}$ is specified in more detail by writing out Gauss' law for $A^c=\nabla_b B^{bc}$ as 
$\int_\Sigma dvn_c A^c = \int_{\partial\Sigma_\infty} da r_b n_c B^{bc} - 
\int_{\partial\Sigma_h} da r_b n_c B^{bc}$
where $n_a$ is the unit normal to $\Sigma$ and $r_b$ is the unit normal to $\partial\Sigma$ within 
$\Sigma$ taken to point towards infinity.   Therefore, we have for the surface volume element $dS_{bc} = 2 da r_{[b}n_{c]}$.  
Note that throughout the paper, we will take $n^a$ to be future pointing.}.
This statement is proved by rewriting the boundary integral as a volume integral using Gauss's law and applying the identity for Killing vectors $\nabla_a\nabla^a  \xi^b = - R^b{}_c\xi^c$.  The resulting volume integrand then vanishes by the vacuum Einstein equations $R_{ab}=0$.  

The Smarr formula for a static black hole
is obtained by  taking the hypersurface $\Sigma$ to extend from the horizon out to spatial infinity.  The boundary of $\Sigma$ then has two components, an inner boundary at the horizon and an outer boundary at infinity.  Equation (\ref{komarrelation}) implies that the difference of the integrals of $\nabla^a\xi^b$ over these two components of the boundary should vanish.
Letting $I_\infty$ and $I_h$ be the integrals of $\nabla^a\xi^b$, multiplied by the normalization factor $(D-2)/8\pi G$, over the boundaries of $\Sigma$ at infinity and the horizon respectively, we have
$I_\infty-I_h = 0$.
It is then straighforward to show that $I_\infty = (D-3)M$ where $M$ is the ADM mass of the black hole, while it was shown in  \cite{Bardeen:1973gs} that the integral at the horizon is given by $I_h=(D-2)\kappa A/8\pi G$.  The Komar integral relation (\ref{komarrelation}) then yields the Smarr formula for static black holes in vacuum Einstein gravity
\begin{equation}\label{smarrvac}
(D-3)M = (D-2) {\kappa A\over 8\pi G}.
\end{equation}
This result agrees with that obtained by means of Euler's theorem and the first law.
The generalization of this result to stationary black holes, which introduces an angular momentum term, comes about in a similar way.  

We now ask whether there is an analogous relation for  black holes in Einstein gravity with $\Lambda\neq 0$?  In this case the field equations imply that the Ricci tensor is given by 
\begin{equation}\label{riccilambda}
R_{ab} ={2 \Lambda\over (D-2)} g_{ab}. 
\end{equation}
A Killing vector then satisfies $\nabla_a\nabla^a  \xi^b = - 2\Lambda/ (D-2) \xi ^b$ and Gauss's law no longer implies the zero on the right hand side of  (\ref{komarrelation}). 
 However, it was shown in references \cite{Bazanski:1990qd,Kastor:2008xb} that a Komar integral relation\footnote{See also reference \cite{Barnich:2004uw} for alternative geometric approach to obtaining a Komar  relation with $\Lambda\neq 0$  and the Smarr formula for AdS black holes.}  that holds with $\Lambda\neq 0$ can be constructed by adding a new term to the integrand in (\ref{komarrelation}).  The new term is built using the anti-symmetric Killing potential $\omega^{ab}$ associated with the Killing vector $\xi^a$ via the relation 
 (\ref{kpot}).  The existence of $\omega^{ab}$ is guaranteed, at least locally, by Poincare's lemma\footnote{This is more familiar in the language of forms.  A Killing vector $\xi^a$ satisfies $\nabla_a\xi^a=0$ as a consequence of Killing's equation.  In terms of the $1$-form $\xi=\xi_adx^a$ this is 
$*d*\xi=0$.  Poincare's lemma then implies that locally $*\xi$ can be written as $*\xi = d *\omega$ where $\omega$ is a $2$-form.  This translates into $\xi = * d* \omega$, or in components $\xi^b=\nabla_a\omega^{ab}$.}.  
The Komar integral relation with $\Lambda\neq 0$ then has the form 
\begin{equation}\label{lambdakomar}
{D-2\over 8\pi G}
\int_{\partial\Sigma}dS_{ab}\left(\nabla^a\xi^b + {2\over D-2}\Lambda\omega^{ab}\right)= 0,
\end{equation}
as one can verify using the Killing identity and the equations of motion (\ref{riccilambda}).  

It is instructive to consider what this new Komar integral relation says for pure Anti-deSitter space.  
In this case,  $\partial\Sigma$ has only one component, at infinity.   Equation (\ref{lambdakomar})  then implies that
the divergent integral of $\nabla^a\xi^b$ is exactly cancelled by a divergence in the integral of the Killing potential.
For a static black hole in AdS, we will see that the divergences at infinity due to the AdS piece and the Killing potential
continue to cancel leaving a finite result.

Before computing these finite pieces, note that 
the Killing potential is not unique.  If $\omega^{ab}$ solves $\xi^b=\nabla_a\omega^{ab}$, then so will
$\omega^\prime{}^{ab} = \omega^{ab} + \lambda^{ab} $, so long as $\nabla_a\lambda^{ab}=0$.
Let us call an antisymmetric tensor $\lambda^{ab}$ that satisfies $\nabla_a\lambda^{ab}=0$ closed, while one that can be written as $\lambda^{ab}= \nabla_c\eta^{cba}$ with $\eta^{cba}$ totally anti-symmetric, and is therefore also divergenceless, will be called exact.  Nonsingular 
antisymmetric $2$-index tensors that are closed, but not exact, in this sense will exist if the rank $D-2$ cohomology of $\Sigma$ is non-trivial.  This will be the case in black hole spacetimes, in which $\Sigma$ is taken to extend outward from the black hole horizon either to spatial infinity for $\Lambda<0$, or to the deSitter horizon for $\Lambda >0$.

We can now ask how this non-uniqueness will effect the Smarr formula that results from the Komar integral 
relation (\ref{lambdakomar}).  Adding an exact term to  the Killing potential will not alter the value of the integral in (\ref{lambdakomar})
on either component of $\partial\Sigma$, since each component is necessarily closed and the integral of a total divergence over a closed surface vanishes.  However, shifting the Killing potential by a term that is closed but not exact will change the values of the integrals on the outer and inner boundaries of $\Sigma$ by equal and opposite amounts.  As a consequence, the integrals of 
$\omega^{ab}$ over the inner and outer boundaries cannot be given separate interpretations.  Only the difference of the two terms is physically meaningful, as is the case with the electromagnetic
potential.

\subsection{Smarr formula for Schwarzschild-AdS black holes}\label{schwarzschildsection}

The route between the Komar integral relation (\ref{lambdakomar}) and the Smarr formula for AdS black holes is more complicated than in the asymptotically flat case.  First, the integrals of $\omega^{ab}$ over the horizon and infinity are tied together by the possibility of adding a closed but not exact term to the Killing potential.  Second, the Killing vector and Killing potential terms in the boundary integrand at infinity are both divergent, in such a way that their divergences cancel to yield a finite result.  In order to see how this works out, we first derive the Smarr formula for the explicit  family of Schwarzschild-AdS spacetimes\footnote{In the asymptotically flat case, the Schwarzschild spacetime has been shown to be the unique static, regular black hole solution (see {\it e.g.} the review \cite{Heusler:1998ua}).  In the asymptotically AdS case, it can be argued that under certain assumptions this is also the case \cite{Anderson:2002xb}\cite{Anderson:2004un}.  However, the situation is considerably more complicated.}.

The Schwarzschild-(A)dS metric 
is  given by 
\begin{equation}\label{schwarz}
ds^2 = - f(r) dt^2 + {dr^2\over f(r)} + r^2 d\Omega_{D-2}^2,\qquad
f(r) = 1 - {\tilde M\over r^{D-3}} -\tl r^2,
\end{equation}
where $\tilde M = 16\pi G M/ (D-2) V_{D-2}$,  $ \tl  = 2\Lambda/(D-1)(D-2)$, $V_{D-2}$ is the volume of a unit $D-2$ sphere, and we will assume that  $\Lambda<0$.  The non-vanishing components of the tensor $\nabla^a\xi^b$ for the static Killing vector $\partial/\partial t$ are given by 
\begin{equation}\label{twist}
\nabla^r\xi^t = -\nabla^t\xi^r ={(D-3) \tilde M\over 2\, r^{D-2}} - \tl r
\end{equation}
The linear term in $r$ leads to a divergent contribution to the boundary integral at infinity in the Komar integral relation (\ref{lambdakomar}).

The Killing potential for the static Killing vector is not uniquely determined.  We will consider the one parameter family of Killing potentials for $\partial/\partial t$, 
\begin{equation}\label{asyomega}
\omega^{rt} = - \omega^{tr} ={ r\over (D-1)} +
 \alpha\,  r_h \left({r_h\over r}\right)^{D-2}
\end{equation}
where $\alpha$ is a dimensionless constant and $r_h$ is the horizon radius\footnote{More generally $\alpha$ could be taken to be a function of the angular coordinates.}.  The linear term in $r$ yields a second divergent contribution to the boundary term at infinity in (\ref{lambdakomar}).
The arbitrary constant $\alpha$ reflects the freedom of adding a closed, but not exact, term to the Killing potential.  
In the case of pure AdS space, the second term in (\ref{asyomega})  is not allowed because of its singularity at $r=0$.
For later use, we take the Killing potential $\omega_{AdS}^{ab}$ for pure AdS spacetime to have non-zero components 
\begin{equation}\label{omegaads}
\omega_{AdS}^{rt} = - \omega_{AdS}^{tr} =r / (D-1).
\end{equation}

Letting $I_h$ and $I_\infty$ again be the components of the integral in (\ref{lambdakomar}) at the horizon and at infinity, we find for $I_\infty$
\begin{equation}\label{inftyint}
I_\infty = - (D-3)M - {2\Lambda V_{D-2}\over 8\pi G}\alpha
\end{equation}
where we have used $dS_{rt}= -(1/2) r^{D-2} d\Omega_{D-2}$.
The divergence coming from  the second term in (\ref{twist})  is exactly cancelled by the divergent term in  the Killing potential (\ref{asyomega}).
The infinite background subtraction of the prescription of \cite{Magnon:1985sc} for computing the Komar mass in asymptotically AdS spacetimes is effectively carried out by cancellations within the boundary integrand itself.
The integral at the horizon is found to be
\begin{equation}\label{horint}
I_h = -(D-2){\kappa A\over 8\pi G} - {2\Lambda V_{D-2} \over 8\pi G} \left(
{r_h^{D-1}\over (D-1) }+ \alpha \right).
\end{equation}

Combining the results (\ref{inftyint}) and (\ref{horint}) using the Komar integral relation $I_\infty - I_h =0$ gives the Smarr formula
\begin{equation}\label{lambdasmarr}
(D-3)M = (D-2) {\kappa\over 8\pi G} A - 2  {\Theta\over 8\pi G}\Lambda
\end{equation}
with $\Theta = -V_{D-2} r_h^{D-1}/  (D-1)$.  This result has the general form (\ref{smarrform}) expected from the scaling argument given in Section (\ref{scaling}) provided we can make the identification 
$\Theta/8\pi G = \partial M/\partial \Lambda$.
We can check that 
this identification is correct by computing $\partial M/\partial \Lambda$ for the Schwarzschild-AdS spacetimes (\ref{schwarz}).
The black hole horizon radius  satisfies $f(r_h)=0$.  From (\ref{schwarz}), we see that this condition implies that
\begin{equation}
M =  {(D-2)V_{D-2}\over 16\pi G} r_h^{D-3} - {V_{D-2}\over 8\pi G(D-1)} r_h^{D-1}\Lambda.
\end{equation}
Taking the derivative of $M$ with respect to $\Lambda$, while holding $r_h$ and therefore the horizon area fixed, reproduces the value of $\Theta$ found above.  This verifies that the geometric derivation of the Smarr formula correctly produces the result expected based on the scaling argument of Section (\ref{scaling}).

It is interesting to note that the value of $\Theta$ above is equal to minus the volume of a $(D-1)$-dimensional ball of radius $r_h$ as computed either in flat space, or equivalently, in anti-deSitter spacetime, using the full $D$-dimensional volume element.   
We will discuss the reason for and physical interpretation of this property in Section (\ref{enthalpy}).

\subsection{More general setting}

In this section we  derive  the Smarr formula in a more general setting, making use only of asymptotic conditions on the metric and properties of the black hole horizon.  Because the Schwarzschild-AdS metric  (\ref{schwarz}) already embodies the general fall-off conditions that we will assume, the calculation required here is quite similar to that  given above.  The benefit will be a general expression, given in equation (\ref{theta}) below, for the quantity $\Theta$ in terms of the Killing potential.

Falloff conditions for asymptotically AdS spacetimes have been discussed in reference \cite{Henneaux:1985tv}.  We will also impose the condition that the angular momentum vanishes and take the asymptotic form of the metric to be $ds^2  \simeq  g_{tt} dt^2 +g_{rr} dr^2 + H\, r^2 d\Omega ^2 _{D-2} $,
with the asymptotic metric functions given by
\begin{equation}\label{asycpts}
g_{tt} =  -\fzero +{c_t \over r^{D-3}} \  , \quad
  g_{rr} =  {1\over \fzero}\left( 1 - {c_r \over \tl r^{D-1}} \right),\qquad
  H = 1 +\tl {c_\theta \over r^{D-1}}
\end{equation}
and $\fzero =1- \tl r^2$.
For the inverse metric, we then have $g^{tt} =   \fzero^{-1}(- 1 + c_t/ \tl r^{D-1} )$
and $g^{rr}= \fzero -c_r / r^{D-3}$.
For solutions to the Einstein equations with $\Lambda<0$, and possibly localized sources of stress 
energy\footnote{
It is straightforward
to include the additional volume integral over matter sources in the Smarr relation,
as was done for $\Lambda =0$ in reference \cite{Bardeen:1973gs}. We allow for such generalizations by phrasing the
current discussion in general terms. In addition, the fall off conditions (\ref{asycpts}) will be needed in
deriving the first law.}, with vanishing angular momentum,
 the asymptotic behavior is simply that of the Schwarzschild-AdS spacetimes and the constants in  
 (\ref{asycpts}) satisfy $c_t =c_r = \tilde M$ and $c_\theta =0$.

The asymptotic behavior of the Killing potential at large radius can similarly be taken to have the form
 (\ref{asyomega}) that it has in Schwarzschild-AdS.
With these asymptotic forms  for the metric and Killing potential, it follows that  the
integrals in the derivation of the  Smarr formula, as well as those in the derivation of  the first law in Section (\ref{firstlawsection}), are finite.  
For a large radius sphere, one then finds for the Killing vector term
\begin{equation}\label{infbt}
da r_b n_c (\nabla^b \xi ^c)  \simeq d\Omega_{D-2} \left( \tl r^{D-1} -{(D-3)\over 2} \tilde M \right).
\end{equation}
We now want to process the Killing potential term at infinity, such that we can leave the form of the Killing potential general, but still provide for the cancelation of divergences.  We do this by both adding and subtracting 
the divergent term $\omega_{AdS}^{ab}$ in (\ref{omegaads}) to and from the Killing potential.  In this way, we are able to write
\begin{equation}
da  r_b n_c ({2\Lambda \over D-2 } \omega ^{bc} ) \simeq -
d\Omega_{D-2}(\tl r^{D-1}) + da  r_b n_c ({2\Lambda \over D-2 } [\omega ^{bc}-
\omega_{AdS}^{bc}] ) 
\end{equation}
We can regard the quantity $\omega^{ab}-\omega_{AdS}^{ab}$ as a renormalization of the Killing potential at infinity.
The subtraction simply removes the (divergent) leading order term in $\omega^{ab}$ near infinity that reflects the asymptotically AdS boundary conditions.
The divergent terms in the boundary integral at infinity then cancel in the same manner as in Schwarzschild-AdS.

For the boundary integral at the horizon, one has
\begin{equation}
I_h = - (D-2) {\kappa A\over 8\pi G} + \int_{\partial\Sigma_h} dS_{ab}\omega^{ab}.
\end{equation}
Combining the boundary integrals according to the Komar integral relation, then again yield the Smarr formula 
(\ref{lambdasmarr}) with $\Theta$ given by
\begin{equation}\label{theta}
\Theta = -\left[ \int_{\partial\Sigma_\infty} dS_{ab}(\omega^{ab}-\omega_{AdS}^{ab}) -
\int_{\partial\Sigma_h} dS_{ab} \omega^{ab}\right].
\end{equation}
This last expression is our general result for $\Theta$.   
We see that $\Theta$ is given by minus the difference between the integral of the renormalized Killing potential at infinity and the integral of the Killing potential on the horizon.

\section{First law with $\delta \Lambda$}\label{firstlawsection}
 In this section we use the techniques of reference  \cite{Sudarsky:1992ty} to derive a version of the first law that includes variations in $\Lambda$.  We will see that the Killing potential again plays an important role and that we once again arrive at the expression (\ref{theta}) for the quantity $\Theta$.
 
 \subsection{Gauss' law from Hamiltonian perturbation theory}
 We start by briefly reviewing the Hamiltonian perturbation theory techniques of references \cite{Traschen:1984bp}
\cite{Sudarsky:1992ty} \cite{Traschen:2001pb}, leading up to the key result in equation (\ref{gaussint}) which allows for the derivation of the first law. Let $\Sigma$
be a family of
 spacelike surfaces\footnote{We use the same symbol to denote both the family of, as well as individual, surfaces.} with  unit timelike normal field $n_a$.
Further,  let $g_{ab}$ be the spacetime metric and $s_{ab}$ the induced metric on $\Sigma$, so that we have 
\begin{equation}\label{metricsplit}
g_{ab} = -n_a n_b +s_{ab} , \qquad n_c n^c =-1 , \qquad n^c s_{cb} =0.
\end{equation}
The Hamiltonian variables are the spatial metric $s_{ab}$ and its conjugate momentum $\pi ^{ab}$. 
Solutions to the Einstein equations with
with energy density $\rho =T_{ab}n^a n^b $ and momentum density $J_a = T_{bc} n^b s^c{}_a$
must satisfy the Hamiltonian and momentum  constraint equations
\begin{equation}\label{constrainteqs}
H=-16\pi G\rho ,\qquad H_a =-16\pi G J_a .
\end{equation}
where $H= -2 G_{ab} n^a n^b$ and $H_a  =-2 G_{bc} n^b s^c{} _a$.
For a cosmological constant stress energy, the constraint equations are simply $H=-2  \Lambda$ and $H_a =0$.

Let $\xi ^a = Fn^a +\beta ^a$ with $n^c \beta _c = 0$ be a vector field.  The Hamiltonian density for evolution along $\xi ^a$ in Einstein gravity with cosmological constant $\Lambda$ is given by 
\begin{equation}\label{hgrav}
{\cal H} = \sqrt{s} \left\{ F (H + 2 \Lambda)+ \beta^a H_a\right\}
\end{equation}
Other sources of stress energy may be included.   See references 
\cite{Traschen:1984bp}\cite{Sudarsky:1992ty} \cite{Traschen:2001pb}
for details of how this affects the result (\ref{gaussint}) below.
Varying the Hamiltonian density (\ref{hgrav}) with respect to $F$ and $\beta ^a$ give the constraint
equations (\ref{constrainteqs}), while
variations with respect to $s_{ab}$ and $\pi ^{ab}$ give the evolution equations for 
$-{\dot \pi} ^{ab}$ and ${\dot s}_{ab}$ respectively, where dot denotes the Lie derivative 
along the vector field $\xi ^a$.

Let $\szero$ and  $\pizero$ be a solution to the Einstein equations with cosmological constant $\lzero$ and 
with  a Killing vector $\xi ^a$. 
Hamiltonian evolution with respect to the 
Killing vector $\xi^a$  implies that  $-{\dot \pizero} =0 ,\  {\dot \szero}=0$. 
Now consider perturbations $s_{ab} = \szero + h_{ab}$,  $\pi ^{ab} = \pizero + p^{ab}$ and $\Lambda =\lzero +\delta\Lambda $ to the spatial metric, the momentum  and the cosmological constant respectively. It then follows from Hamilton's equations
 for the zeroth order spacetime, that the linearized constraint  
 operators $\delta H$ and $\delta H_a$ combine to form a total derivative, according to
 \begin{equation}\label{icvid}
 F\delta H + \beta ^a \delta H_a =- D_c B^c
 \end{equation}
 where $D_a$ is the covariant derivative operator on $\Sigma$ compatible with the metric $\szero$, and the spatial vector $B^a$ is given by 
 \begin{equation}\label{boundaryterm}
B^a=F({D}^ah-{D}_bh^{ab})-h{D}^aF+h^{ab}{D}_bF+
{1 \over \sqrt{{s}}}\beta^b({\pi_{(0)}}^{cd}h_{cd}{s}^{(0)a}{}_b-2{\pi}_{(0)}^{ac}h_{bc}-2p^a{}_b)
\end{equation}
 This last statement holds for arbitrary perturbations $h_{ab}$, $p^{ab}$ and $ \delta\Lambda$. 
If the perturbations
 are taken to be solutions to the linearized Einstein equations, equation (\ref{icvid}) implies that  the linearized constraints
 (\ref{constrainteqs}) take the form of a Gauss' Law\footnote{Including general perturbative matter sources would give an additional contribution  on the right hand side of the form
 $16\pi G(F\delta \rho +\beta ^a \delta J_a) $,
which would act as a source for the geometrical fields on the left hand side.} 
 \begin{equation}\label{gauss}
 D_c B^c =  2F\delta \Lambda.
 \end{equation}
 
 The  cosmological constant source term in equation (\ref{gauss}) 
 can be rewritten as a total derivative by once again making use of the Killing potential $\omega^{ab}$.
 One has the chain of equalities $F=-n_a \xi ^a = -D _c( n_a \omega ^{ca} )$.  Substituting
 into  equation (\ref{gauss}) and rewriting it  in integral form  then 
\begin{equation}\label{gaussint}
\int _{\partial \Sigma }
 da_c (B^c  +2 \omega^{cd}n_d\delta\Lambda ) =0.
\end{equation}
 Equation (\ref{gaussint}) is the main result of this summary.  For the familiar case of
 a static black hole in an asymptotically flat spacetime, {\it i.e.} with $\lzero=\delta\Lambda =0$, evaluating the boundary term  on the boundaries at the  horizon and at infinity gives the usual first law \cite{Sudarsky:1992ty}.  It is worth noting that this proof of the first law holds under less stringent assumptions than that given in \cite{Bardeen:1973gs} in that the perturbations are not required to share the static symmetry of the background black hole spacetime.

\subsection{First law}

 In this section we derive the first law including changes in the cosmological
 constant $\delta\Lambda$ by
  evaluating the boundary terms in (\ref{gaussint}) when $ \gzero$ is  a static, asymptotically
 AdS black hole with bifurcate Killing horizon. Consider perturbations about the metric $g^{(0)} _{ab}$, with the perturbations
also required to satisfy the asymptotically AdS fall-off conditions (\ref{asycpts}).
Choose the Killing vector  $\xi^{a}$  in the Hamiltonian construction of the previous subsection 
 to be the horizon generator and assume that $\xi^a$ approaches
 $ (\partial/\partial t)^{a}$ at infinity in the asymptotic coordinates used above.
The spacelike hypersurface 
 $\Sigma$ in the Gauss' law construction is taken to extend from a boundary $\partial \Sigma_{h}$ at the bifurcation sphere of the black hole horizon  
to a boundary $\partial \Sigma_{\infty}$ infinity, chosen 
  such that at infinity the unit normal is $n_a = -F \nabla_a t$.  With these choices, the terms proportional to the vector $\beta^a$ in the boundary term (\ref{boundaryterm}) vanish sufficiently rapidly at infinity that they do not contribute to the boundary term there.

Following our practice in section (\ref{smarrsection}), let us write equation (\ref{gaussint}) as
$I_{\infty}-I_{h}=0$,
where again the normal to $\partial\Sigma_h$ within $\Sigma$ is taken to point outward towards infinity.
First consider the boundary term at infinity and compute the integral
\begin{eqnarray}
\int _{\partial \Sigma_{\infty}} da_cB^c=
\int _{\partial \Sigma_{\infty}} da_c [F(D^{c}h-D_{b}h^{bc})-hD^{c}F+h^{bc}D_{b}F]
\end{eqnarray}
At large radius it is sufficient to consider both the background metric and the perturbations to have the Schwarzschild-AdS form (\ref{schwarz}).  Near infinity, we then have
\begin{eqnarray}
h_{rr}\simeq {-1\over f^2}\delta f 
=- \frac{1}{f(r)^{2}} \left[\frac{\delta \tilde M}{r^{D-3}} + \delta\tilde\Lambda r^2\right],
\end{eqnarray}
while $F\simeq \sqrt{f}$ and $da_r\simeq r^{D-2}d\Omega_{D-2}/\sqrt{f}$ and we find that
\begin{equation} \label{divergingtermone}
\int _{\partial \Sigma_{\infty}} da_cB^c = -16\pi G\delta M-\lim_{r \to \infty} \left(\frac{2 r^{D-1}V_{D-2}}{D-1}\right)\delta \Lambda.
\end{equation}
Evaluating the second term in the boundary integral at infinity using the asymptotic form of 
$\omega^{ab}$  in (\ref{asyomega}) leads to
\begin{equation}\label{divergingtermtwo}
2\int _{\partial \Sigma_{\infty}} da_c  \omega^{cd} n_d \delta\Lambda =
 \lim_{r \to \infty}\left( \frac{2 r^{D-1}V_{D-2}}{D-1}\right) \delta \Lambda + 
 2\left( \int_{\partial\Sigma_\infty} da_c(\omega^{cd}-\omega_{AdS}^{cd}) n_d\right)\delta \Lambda.
\end{equation}
Combining the results in equations (\ref{divergingtermone}) and (\ref{divergingtermtwo}) to obtain 
$I_\infty$, we see that the divergent pieces cancel, much as they did in the derivation of the Smarr formula above.  The remaining finite expression for the boundary term at infinity is given by 
\begin{equation}
I_{\infty}=-16 \pi \delta \mass + 2\left( \int_{\partial\Sigma_\infty} da_c(\omega^{cd}-\omega_{AdS}^{cd}) n_d\right)\delta \Lambda.
\end{equation}
Now, consider the boundary term in (\ref{gaussint}) at the horizon. Evaluation of the first term in the integral  proceeds as in references 
\cite{Sudarsky:1992ty,Traschen:2001pb} and gives $-2\kappa \delta A$.  
One then has 
\begin{equation}
I_h = -2\kappa \delta A + 2\left(\int_{\partial\Sigma_h} da_c\omega^{cd}n_d\right)\delta\Lambda
\end{equation}
Putting all the terms together gives the first law with varying cosmological constant
\begin{eqnarray}\label{finalfirstlaw}
\delta M=\frac{\kappa}{8\pi G}\delta A+ \frac{\Theta}{8\pi G}\delta \Lambda
\end{eqnarray}
with $\Theta$ again given by the expression in (\ref{theta}).   In the next section we will show that the  new $\Theta\delta\Lambda$ term in the first law for AdS black holes can be interpreted as a work term of the form $VdP$.

\section{Enthalpy}\label{enthalpy}

The quantity $\Theta$ is expressed in (\ref{theta}) in terms of the boundary integrals of the renormalized Killing potential at infinity and the Killing potential at the horizon.  
We can gain physical insight into the thermodynamic role of $\Theta$ by re-expressing
 these boundary integrals as volume integrals.

Suppose that the hypersurface $\Sigma$ is orthogonal to the static Killing vector $\xi^a$, so that  $\xi ^a = Fn^a$.
The full spacetime volume element on $\Sigma$ is given by $\sqrt{-g^{(D)}}= F \sqrt{g^{(D-1)}}$, where $\sqrt{g^{(D-1)}}$ is the
intrinsic volume element on $\Sigma$.
One of the ingredients of  the expression for $\Theta$ in (\ref{theta}),
the integral of $\omega^{ab}$ over the boundary of $\Sigma$
 can then be rewritten via the chain of equalities
\begin{equation}
\int_{\partial\Sigma} dS_{ab}\omega^{ab} 
= \int_\Sigma d^{D-1}x \sqrt{g^{(D-1)}} n_b\xi^b 
= -\int_\Sigma d^{D-1}x \sqrt{-g^{(D)}}.
\end{equation}
as minus the quantity $V_{BH} \equiv \int_\Sigma d^{D-1}x \sqrt{-g^{(D)}}$, the (infinite) volume on $\Sigma$ between the black hole horizon and infinity. The other ingredient in $\Theta$, the integral of $\omega^{ab}_{AdS}$ over the boundary at infinity can be similarly written as $V_{AdS}$, the (also infinite) volume of a spatial slice of AdS spacetime stretching out to infinity.
We then have the result
\begin{equation}\label{subtractvols}
\Theta = V_{BH}-V_{AdS}.
\end{equation}
At the end of Section (\ref{schwarzschildsection}), we observed that the value of $\Theta$ was equal to minus the volume of a ball of radius $r_h$ in AdS spacetime.  Here we see the origin for this result.  The quantity $V=-\Theta $ gives a measure of the volume excluded from the spacetime by the black hole horizon\footnote{The Smarr formula for a family of deSitter black holes is computed in \cite{Brihaye:2008br} by integrating over the region between the black hole and deSitter horizons including the stress energy volume contribution directly, producing a term proportional to the volume between the horizons. }.

We are now in a position to give a physical interpretation of the term $\Theta\delta\Lambda/8\pi G$ in the first law 
(\ref{finalfirstlaw}).  The cosmological constant can be thought of as a perfect fluid stress-energy with pressure 
$P = -\Lambda/8\pi G$.  With these identifications, we have $\Theta\delta\Lambda/ 8\pi G = V\delta P$.
Combining this with the usual identifications $T=\kappa/2\pi$ and $S = A/4G$ for the black hole temperature and entropy, the first law becomes
\begin{equation}
\delta M = T \delta S + V\delta P.
\end{equation}
Stated in this way, the right hand side coincides with the variation of the enthalpy $H= E+PV$ in classical thermodynamics.
This result suggests that the mass of an AdS black hole should be thought of as the enthalpy of the spacetime.

The identification of the AdS black hole mass as an enthalpy makes good physical sense. 
The mass of an AdS black hole is defined via an integral at infinity.  However, between the black hole horizon and infinity is an infinite amount of energy density that needs to be subtracted off in some manner to get a finite result.  Since the energy density in the cosmological constant is $\rho= + \Lambda/8\pi G$, adding a $PV$ term naturally cancels out a $\rho V$ contribution to the energy.  

Specific constructions in which such a subtraction takes place are the following.
 In the spinorial proof of the positive mass theorem for asymptotically AdS spacetimes \cite{Gibbons:1982jg}, 
 a super-covariant derivative operator is introduced, together with the super-covariantly
 constant spinors which exist in pure AdS. This derivative operator has the effect of
 subtracting out the contribution of the cosmological constant in the positve energy construction. 
In effect, the background pressure times the volume is
 added to the contributions to the mass from the matter and gravitational fields. 
 The Abbott-Deser construction for conserved
 charges in AdS \cite{Abbott:1981ff} has the same feature of adding in the AdS $PV$ to give a finite
 conserved charge.  It would further be interesting to study whether the enthalpy interpretation of the AdS black hole mass 
 proves useful from the standpoint of Euclidean black hole thermodynamics.

\section{ Killing potential near a static horizon}\label{killingpotential}
The potential for the Killing vector $\xi ^a$
was defined as an antisymmetric tensor $\wab$ whose divergence gives $\xi ^a$ as in equation (\ref{kpot}).
 In this section we show that there exist solutions for $\wab$ that are well behaved 
 near the horizon. These near horizon solutions take the form of the area element on the horizon times
 an arbitrary  function $\psi$  that is Lie-derived by $\xi ^a$.
 The Killing potential is not unique, as a solution to the homogeneous equation can always be
 added in, and this is reflected in the arbitrariness of $\psi$. However, only the constant
 mode $w_H$ in $\psi$ contributes  to the boundary integral of $\wab$ over the horizon. 
 Further, a homogeneous solution with a different tensor structure also does
 not contribute to the boundary integral. Hence the horizon boundary term has an arbitrariness
 of a single constant. Similar remarks apply to the integral of $\wab$ at infinity. 
 
Physically
 meaningful statements, such as the Smarr relation or the first law, depend on the difference
 in the potential between the horizon and infinity. This  is  analogous
 to the situation with the electric potential, in which only the difference in the electric potential times
 $\delta Q$ contributes to the work term in the first law. In the electromagnetic case,
 this work term can be understood in terms of 
 the mechanics of charged particles. For the Killing potential term in (\ref{finalfirstlaw}) we still lack such a 
 test particle (or test field) based mechanical understanding.
 
 Before turning to the construction of the near horizon solutions,
  we note two different gauge choices for the potential that
 give alternate interpretations of $\Theta$, which is given in (\ref{subtractvols}) as the
  finite difference between the volume of a spatial slice in pure AdS and for an AdS  black hole. 
 As we saw explicitly in the case of Schwarzschild-AdS, this gives the picture that $\Theta$ is (the negative of) the volume occupied black hole at
  a given time.
   Choose a gauge so that
  $\wab$ is equal to its AdS value at infinity, then $\Theta$ is given in (\ref{theta}) by just
  the integral of $\wab$ over the horizon. The results of this section then show that this is equal
  to $-w_H A_H $. So the meaning of the constant value of $\wab$ on the horizon is the
  length so that $w_H$ times the horizon area is an effective volume of the black hole.  Alternatively, one can choose a gauge such that the Killing potential vanishes on the horizon.
  
\subsection{Preliminaries} 
We start by collecting some needed information about the geometry in a neighborhood of
a Killing horizon $\hor$ with null normal $\xi^a$. Here we follow the treatments of Carter 
\cite{Carter:1973} \cite{Carter:1987hk} and  Price and Thorne \cite{Price:1986yy}. Since
 $\xi^a$ is also tangent to $\hor$ we can choose a coordinate $s$ such that
 \begin{equation}\label{xiinfo}
 \xi^a =\left({\partial \over \partial s}\right)^a .
 \end{equation}  
Now, let $x^i$ with $i=1,\dots,D-2$ denote the $D-2$ spatial coordinates on a constant $s$ cross section of the horizon.
The vectors 
 $\ei ^a    =({\partial \over \partial x^i }){}^a$ 
  satisfy $\xi _a \ei ^a =0$. 
 
In order to extend these coordinates to a neighborhood of the horizon, we 
take the additional coordinate $U$ to be an affine parameter for the null geodesic field $q^a$ that
is ingoing at $\hor$, orthogonal to the vectors $\ei ^a$ and normalized so that $ q^a \xi_a  =-1$ on
 $\hor$.
Hence, we have
\begin{equation}\label{qinfo}
q^a =\left({\partial\over \partial U}\right)^a,\qquad    q_a q^a =0,\qquad  q^b \nabla _b q^a =0,\qquad   \ei ^a q_a |_{\hor} =0.
\end{equation}
The horizon is taken to be at $U=0$.
The horizon surface gravity is  given in terms of $\xi^a $and $q^a$  by 
\begin{equation}\label{surfgrav}
\kappa = \xi ^b q^c \nabla _c \xi _b ={1\over 2} q^c \nabla _c (\xi \cdot \xi ),
\end{equation}
and because the components of the vorticity of the Killing field, $\nabla _{[a} \xi _{b]} =\nabla _a \xi _b$, vanish when projected onto the horizon, it follows from equation (\ref{surfgrav})  that the vorticity can be written as
\begin{equation}\label{horvort}
\nabla _a \xi _b = \kappa ( \xi _a q_b - q_a \xi_b).
\end{equation}

The coordinates $s$ and $x^i$ can now be extended off the horizon by parallel transporting the
vectors $({\partial \over \partial s} )^a$ and $ ({\partial \over \partial x^i })^a$ outward from the horizon along $q^a$. In the resulting coordinates, one has $g_{UU} =0$, because $U$ is a null coordinate throughout, while on the horizon $U=0$, one has $g_{ss}=0 $. 
Near the horizon, the metric then has the form  \cite{Price:1986yy} 
\begin{equation}\label{nearhormetric}
ds^2 = -2ds dU +2K U(ds)^2 +\gamma _{ij} dx^i dx^j +{\cal O}(U^2 ),
\end{equation}
where
the metric functions $\kappa$ and $\gamma_{ij}$ may depend on the coordinates $s$ and $ x^i$ (although on the horizon they are independent of $s$ which  is a Killing parameter there). It is straightforward to verify that
with this form of the metric, $q^a$ indeed solves the geodesic equation and that the coordinate basis vectors 
$({\partial \over \partial s} )^a$ and $ ({\partial \over \partial x^i })^a$
are parallel transported by $q^a$.

In order to evaluate derivatives of $\xi^a$ in a direction that is not tangent to $\hor$,
we need to know how the Killing field extends off $\hor$ in the coordinates of (\ref{nearhormetric}).
Define the vector $l^a=({ \partial \over \partial s})^a $ in a neighborhood of $\hor$.  The vectors 
 $l^a$ and $\xi^a$ will then agree on $\hor$, but otherwise may differ.  However as we next show, the vector $l^a$
 also satisfies Killings equation on $\hor$.  In coordinates of equation (\ref{nearhormetric}), one has $l_b =g_{bs}$, and 
 \begin{eqnarray}
 q^a l^b \nabla _{(a} l_{b)} & =&  \partial _U g_{ss} +\partial _s g_{Us}  -2\Gamma ^c _{sU} g_{cs}
  \\  \nonumber
  &=& -2\kappa +2\kappa =0
  \end{eqnarray}
  Similarly, one can check that 
   $\ei ^a q^b (\nabla _a l_b + \nabla _b l_a )=0$
  and  $q^a q^b \nabla _a l_b = \partial _U g_{Us}
  -2\Gamma ^c _{UU} g_{cs}=0$   on $\hor$.
   Hence we can use $({ \partial \over \partial s}) $ for the Killing field and its derivatives
   on $\hor$.   Finally, the surface gravity is given by
  $ q^a \nabla _a (\xi_b \xi ^b) = \partial _U (g_{ss}) =2K$, so that the metric function $K$ evaluated at the horizon is given by the surface gravity, {\it i.e.}
   \begin{equation}\label{kk}
   K|_{\hor} =\kappa  .
   \end{equation}
   
\subsection{ The Killing potential}
We look for a solution to  (\ref{kpot}) for $\wab$  in a neighborhood of the horizon $\hor$ that has the form
\begin{equation}\label{kpotguess}
\wab =\psi ( \xi ^a q^b - q^a \xi^b)
\end{equation}
for some function $\psi$. As previously mentioned, this is the only index structure for
$\wab$ that will contribute to the integral over the horizon.
 By \textit{in a neighborhood} of $\hor$, we mean such that the resulting expression for 
$\wab$ correctly 
reproduces the Killing vector $\xi ^a$ and its derivatives on the horizon, so that Killing's equation is satisfied on $\hor$.
If $\omega^{ab}$  is to be a Killing potential, then on $\hor$  it must reproduce the conditions 
$\xi_a\xi^a=0$ and $\xi^aq_b=-1$.   These translate respectively into the conditions
\begin{equation}\label{psicond}
\xi^b \nabla _b \psi =0 ,\quad  \nabla_b (q^b \psi ) =-1.
\end{equation}
 Lastly, we need that $\xi _a \ei ^a =0$ . If the vectors $\ei ^a$ are chosen such that
they commute with $\xi^a$, as in the previous sub-section, this orthogonality condition is
also satisfied.

The form (\ref{kpotguess}) must also reproduce Killing's equation on the horizon. We have
already shown that $ ({\partial \over \partial s}) $ satisfies Killing's equation on $\hor$.  Therefore,
we just need to show that $\nabla_ a \wab = ({\partial \over \partial s})^b $ plus terms
of order $U^2$. For example, the $b=s$ component should give $\xi ^ s =1$. This is checked by computing
\begin{eqnarray}\label{checks}
\nabla _a \omega ^{as} &=&
 {-1\over \sqrt{\gamma} } \partial _U (\sqrt{\gamma} \psi ) (1+ {\cal O}(U^2 )) \\ \nonumber
 &=& (1+ {\cal O}(U^2 ))
\end{eqnarray}
where we have used the fact that $\sqrt{-g} =\sqrt{\gamma } (1+ {\cal O}(U^2 ))$ in the first line and equation
(\ref{psicond}) in the second. Similarly, one finds that $\nabla _a \omega ^{aU} =
\nabla _a \omega ^{ai} =0$ plus terms of order $U^2$.

To solve for $\psi$ we Taylor expand  about the horizon,
\begin{equation}\label{taylor}
\psi (U, s, x^i ) = \psi _0 + U\partial _U \psi |_{\hor} + {1\over 2}U^2 \partial _U ^2 \psi |_{\hor}
+...
\end{equation}
The first equation of (\ref{psicond}) becomes $\partial _s \psi _0 =0$, and the second
says that $\partial _U \psi |_{\hor} =-1$. Hence
\begin{equation}\label{psisol}
\omega ^{sU} = \psi = \psi _0 (x^i ) -U +{\cal O} (U^2 )
\end{equation}
for the only nontrivial component of $\wab $.

As a check, let's compute the one non-zero antisymmetric derivative of the Killing field on the horizon equation (\ref{surfgrav}). Substituting $\xi_b =g_{bs}$,
we want to check that at $U=0$
\begin{equation}\label{kappacheck}
2\kappa= {\partial \over \partial U} \left( {1\over \sqrt{-g}}g_{bs} \partial _c ( \sqrt{-g} \omega ^{cb})
\right)
\end{equation}
Using $\xi = ({\partial \over \partial s})$ and $ \sqrt{-g} =\sqrt{\gamma } $
to order $U^2$, and $\partial _U \psi =-1$ at $U=0$,
this becomes $2\kappa =\partial _U g_{ss} -\partial_s \partial _U \psi=
2\kappa +0$.

Therefore the general solution for $\wab$ on a Killing horizon with the tensor structure 
 (\ref{kpotguess}) is
 given by  (\ref{psisol}). We have not used the field equations,
  so this will be true in the higher derivative Lovelock theories as well as
 in Einstein gravity with cosmological constant. There is an
 arbitrary function $\psi _0$, which can depend on the spatial coordinates on the horizon,
 which  is just the gauge freedom in the definition of the potential. This means that
 the potential can always be chosen to be constant on the horizon intersected
 with a spacelike slice, since one can always add in some function of the $x^i$ to
 cancel whatever position dependence of $\psi$ is present. Further, if $\psi$ is expanded
 in an orthonormal set of basis functions on the horizon, only the constant mode contributes
 to the boundary integral.

\subsection*{Acknowledgements}

The authors thank Piotr Chrusciel, Luca Grisa, Julio Oliva, Lorenzo Sorbo, Ricardo Troncoso and Jorge Zanelli for helpful discussions and correspondence.  
The work of DK and JT is supported by NSF grant PHY-0555304.
This work of SR was partially funded by FONDECYT grant 3095018.
Centro de Estudios Cient'ficos (CECS) is funded by the Chilean
Government through the Millennium Science Initiative and the Centers of
Excellence Base Financing Program of CONICYT. CECS is also supported by
a group of private companies which at present includes Antofagasta
Minerals, Arauco, Empresas CMPC, Indura, Naviera Ultragas and Telef—nica
del Sur. CIN is funded by CONICYT and the Gobierno Regional de Los R'os.

\end{document}